\documentclass[sigconf]{acmart}
\copyrightyear{2020} 
\acmYear{2020} 
\setcopyright{acmcopyright}\acmConference[ICPC '20]{28th International Conference on Program Comprehension}{October 5--6, 2020}{Seoul, Republic of Korea}
\acmBooktitle{28th International Conference on Program Comprehension (ICPC '20), October 5--6, 2020, Seoul, Republic of Korea}
\acmPrice{15.00}
\acmDOI{10.1145/3387904.3389294}
\acmISBN{978-1-4503-7958-8/20/05}

\usepackage{ragged2e}
\usepackage{listings, lstpresto}
\newcommand{\pseudo}[1]{\small{}Data Model for~\emph{#1}}

\newcommand{\defwip}{i) engineers rely on a multitude (perhaps
  hundreds) of loosely integrated tools; ii) engineers engage in
  concurrent and relatively long running workflows; ii) infrastructure
  (such as logging) is not fully aware of work items; iv) engineering
  processes (e.g., for incident response) are not explicitly modeled.}

\newcommand{\myabstract}{Understanding what a software engineer (a
  developer, an incident responder, a production engineer, etc.) is
  working on is a challenging problem~--~especially when considering
  the more complex software engineering workflows in
  software-intensive organizations: \defwip{} In this paper, we
  explain the corresponding `\emph{work-item prediction challenge}' on
  the grounds of representative scenarios, report on related efforts
  at Facebook, discuss some lessons learned, and review related work
  to call to arms to leverage, advance, and combine techniques from
  program comprehension, mining software repositories, process mining,
  and machine learning.}

\newcommand{\mykeywords}{developer workflow, loose tool integration,
  concurrent workflow, process mining, machine learning, code
  similarity, word correlation}

\definecolor{codebackground}{rgb}{0.97,0.97,0.97}
\definecolor{uribackground}{rgb}{0.9,0.9,0.9}
\definecolor{gray}{rgb}{0.3,0.3,0.3}
\definecolor{keyword}{rgb}{0.5,0.0,0.5}
\definecolor{comment}{rgb}{0.3,0.5,0.3}
\definecolor{code}{rgb}{0.0,0.0,0.3}
\definecolor{ncode}{rgb}{0.0,0.3,0.3}
\definecolor{scode}{rgb}{0.3,0.0,0.3}
\definecolor{static}{rgb}{0.0,0.0,0.3}

\setcopyright{none}

\begin{document}

\title{Understanding What Software Engineers Are Working on\vspace{50\in}}
\subtitle{The Work-Item Prediction Challenge\vspace{50\in}}

\author{Ralf L\"ammel, Alvin Kerber, and Liane Praza}
\authornote{This paper appears in
Proceedings of 28th International Conference on Program Comprehension,
ICPC 2020. The subject of the paper is
  covered by the first author's keynote at the same conference.}
\affiliation{Facebook Inc.\vspace{150\in}}

\begin{abstract}
\myabstract\vspace{150\in}
\end{abstract}

\keywords{\mykeywords\vspace{150\in}}

\maketitle


\section{Introduction}

A `\emph{process-unaware (information) system}'~\cite{GoelBW13} does not expose processes in a direct manner at an architectural and user level. In this paper, we are concerned with a very similar problem in the context of the ecosystems and processes that engineers use to develop, to deploy, and to maintain software systems: the challenge of predicting what work item a software engineer is working on:

\begin{center}\itshape{}The work-item prediction challenge\end{center}

\newpage

\begin{figure}[t!]
{

\centering

\includegraphics[width=.9\columnwidth]{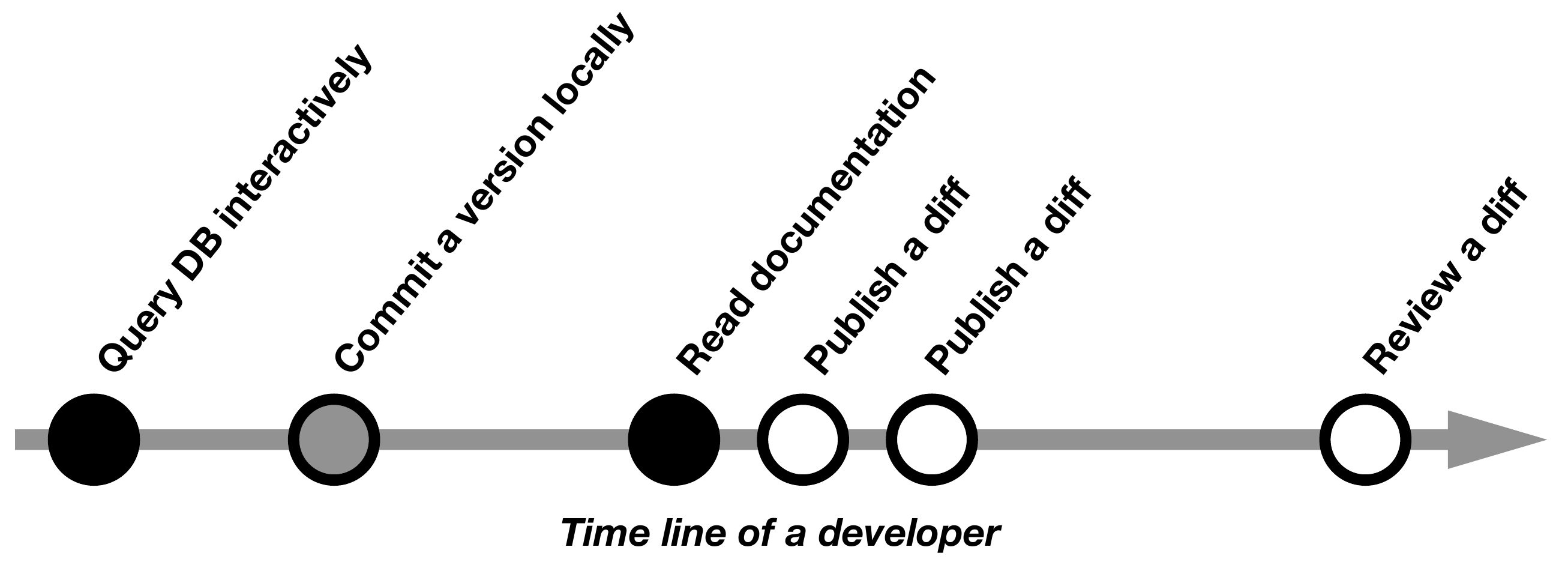}

}

\vspace{-22\in}

\justify\small
The events on the timeline concern different `diffs' (i.e., system changes all the way from committing a change locally to landing the change in production) as work items. White events are trivially associated with diffs. Gray events require dedicated data integration for association. Black events are hard to associate; advanced heuristics and machine learning may be of use. That is, which of the three diffs should be associated with the DB query and the documentation access?

\vspace{-42\in}
\caption{Dark Matter in Engineering Workflows.}
\label{F:dark-matter}
\vspace{-32\in}
\end{figure}

\noindent
Abbreviation: WIP challenge.\footnote{WIP tends to serve also as an acronym for `work in progress', which is very fitting for our purposes because predicting what work item is being worked on essentially boils down to tracking all work in progress, at all times, as we will discuss in more detail.} We speak of a challenge here because of these \textbf{defining characteristics}: \defwip{} In combination, these characteristics give rise to what we call `\emph{dark matter}'; see \autoref{F:dark-matter} for an illustration.


Being able to predict the work item along the timeline of each developer has profound applications, for example, in the context of incident response in engineering (with relevance for reliability, integrity, privacy, and security) or the aggregation of key performance indicators for engineering processes.

\paragraph*{\textbf{Call to Arms}} While there exists significant related work on capturing and analyzing workflows of software engineers (e.g., in terms of the use of VCSs or IDEs~\cite{Mockus03,PoncinSB11,IoannouBW18}), this paper calls to arms on research addressing the WIP challenge in terms of the defining characteristics to enable work-item prediction for software engineering workflows in software-intensive organizations. Future work is needed to leverage, advance, and combine techniques from program comprehension, mining software repositories, process mining, and machine learning.

\paragraph*{\textbf{Roadmap of the Paper}} We explain the WIP challenge in more detail on the grounds of representative scenarios (Section~\ref{S:scenarios}), report on related efforts at Facebook (Section~\ref{S:facebook}), discuss some lessons learned (Section~\ref{S:lessons}), and review related work (Section~\ref{S:related}).


\pagebreak


\begin{figure*}[t!]
\centering
\includegraphics[width=.82\textwidth]{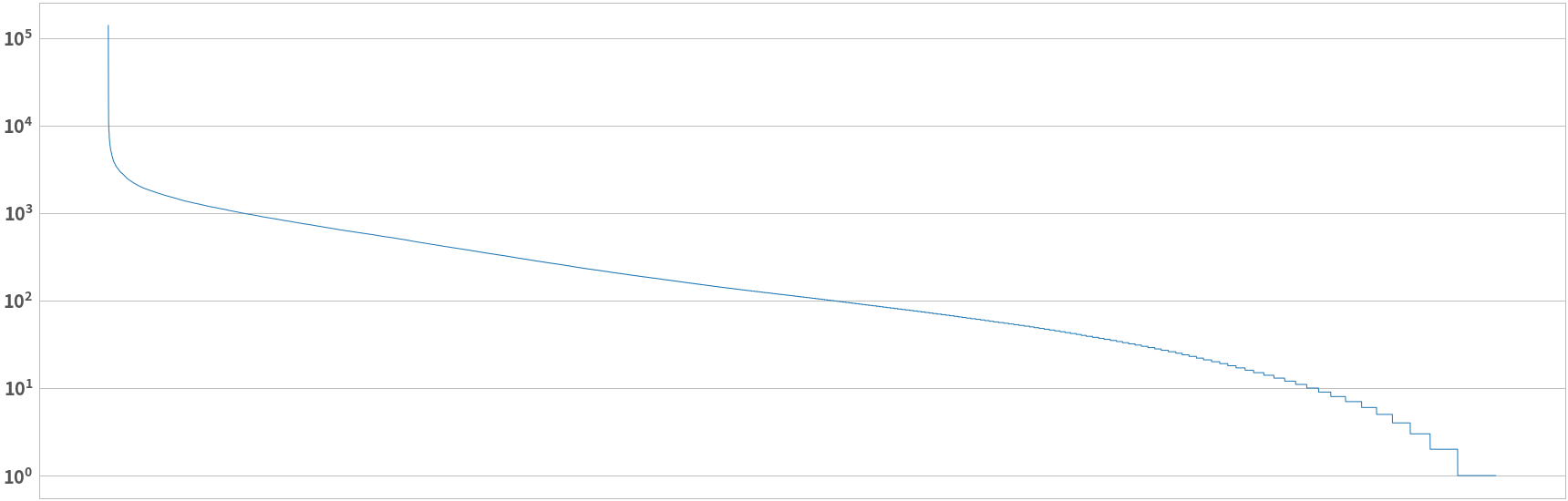}
\vspace{-22\in}
\caption{Number of (selected) tools used per employee on a given day for many of Facebook's employees.}
\label{F:tools-per-employee}
\vspace{22\in}
\end{figure*}


\section{Representative Scenarios}
\label{S:scenarios}

At Facebook, we have encountered the need for and engaged in efforts towards addressing the WIP challenge on several occasions. We now describe two different scenarios in which work-item prediction is required. We inject some data points or illustrations to underline the defining characteristics of the WIP challenge. For non-disclosure reasons, we cannot report more specifics of any actual WIP scenarios at Facebook, but we consider the selected scenarios as representative of what is needed in software-intensive organizations.


\subsection{The `Incident Response' Scenario}

\paragraph*{Summary} Engineers need to respond to an incident (an alert) such as suboptimal performance of an important system component. To this end, engineers would like to reuse workflow steps by experienced engineers who investigated and mitigated similar incidents in the past.

\paragraph*{Details} \mbox{}

\begin{itemize}
  
\item For some types of incidents, there may exist team-specific or more generic documentation with workflows for incident response, but it may be outdated or too unspecific. In fact, there are so many different kinds of incidents and the response workflows change over time. Thus, some form of `automatic documentation' is needed.

\item If we were to extract workflow steps from past incident responses, we need to identify all events associated with a given incident ID. Available logging does not suffice for such association in practice, due to a multitude of loosely integrated tools. Simple heuristics are insufficient because
engineers engage in concurrent and relatively long running workflows.

\item We may attempt reverse engineering and data mining to recover the incident IDs from logs. This would be a continuous and possibly prohibitively expensive effort, given the complexity and the evolution of the tool suite to be considered.

\item We may instead attempt re-engineering to improve, for example, logging, thereby improving tool integration. This would risk reliability of the infrastructure / the ecosystem. This would also cause disruption, as the integration would affect engineering processes, which would also be the case, if we were to start from the premise of explicit modeling of engineering processes.

\end{itemize}

\paragraph*{Illustration~--~Multitude of Tools}

\autoref{F:tools-per-employee} illustrates tool usage at Facebook. The chart shows the number of tools used per employee per day. The employees with more tools used per day are mostly developers and other engineers. We note that the counting scheme for tools is idiosyncratic; we often treat tool suites (sometimes of significant size) as single tools because it is easier for us to count that way; we also filter tools in some ways.


\begin{figure*}[t!]
\parbox{.74\textwidth}{
  \includegraphics[width=.739\textwidth]{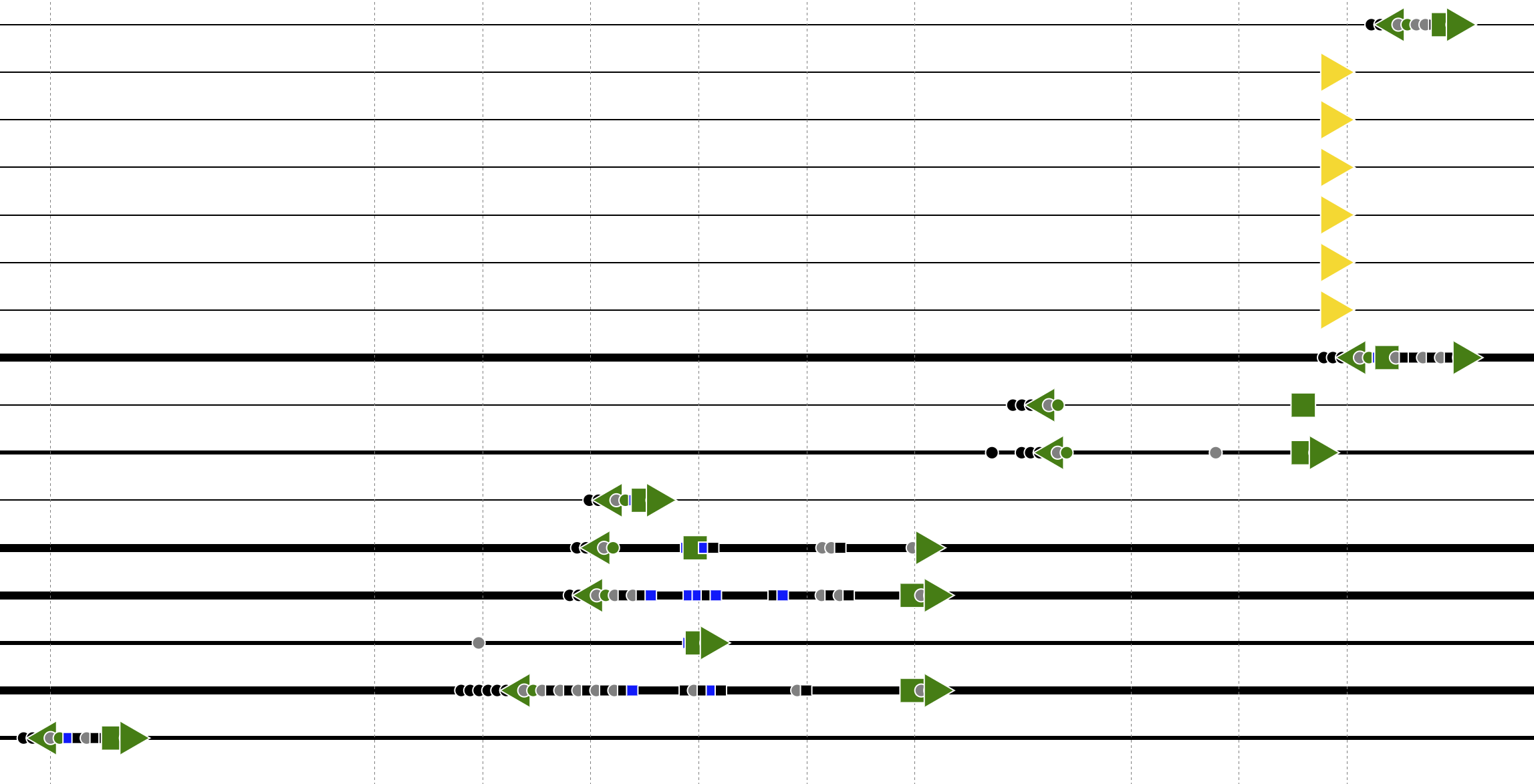}
}\hfill\parbox{.24\textwidth}{
\includegraphics[width=.239\textwidth]{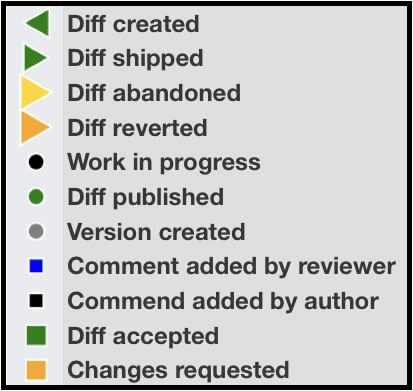}
}
\vspace{-33\in}
\caption{Concurrent workflow by a developer on several diffs (y-axis) over a few days (x-axis).}
\label{F:diff-analyzer}
\vspace{22\in}
\end{figure*} 


\subsection{The `Aggregate Performance' Scenario}
\label{S:KPI}

\paragraph*{Summary} Organizations aggregate performance data at different levels by means of Key Performance Indicators (KPIs) to inform decision making, for example, regarding the effectiveness of engineering practices; see~\cite{JaspanS19} for some broader context on productivity metrics. Let us consider two KPIs related to `diffs'  (i.e., system changes all the way from committing a change locally to landing the change in production) as work items: i) `time spent on reviewing a diff'; ii)`time spent on a diff in total'. (In practice, these KPIs would be set up in a nuanced manner, for example, by grouping by product area.)

\paragraph*{Details} \mbox{}

\begin{itemize}
  
\item Consider the KPI `time spent on reviewing a diff' first. This KPI is still implementable, if we assume that we can leverage logging for individual review comments, review decisions, and engagement with the reviewing UI (such as clicking, typing, and scrolling).

\item Now consider the more general KPI `time spent on a diff in total'. Logging may not cover important slices of work on a diff during 'idea formation' (e.g., reading documentation) or extra investigations independent of the diff's code (e.g., interactive database queries).

\item Concurrency of working on diffs is entirely to be expected, especially in keeping several published diffs active for reviews and revisions until eventually shipping them independently or together. These workflow are also potentially long-running; think of diff authors and reviewers residing in different time zones.

\end{itemize}

\paragraph*{Illustration~--~Concurrent and Long-Running Workflows}

\autoref{F:diff-analyzer} illustrates diff-related workflows of a developer over a few days. It is clear that the developer works on several diffs concurrently, as indicated by the overlapping distances between `diff created' and `diff shipped' / `diff abandoned' events. In some cases, key events of several diffs occur together on the timeline, but it is reasonable to assume that some `context switching' must have occurred prior to these `checkpoints'. Note also the `work-in-progress' events which appear on the timeline at a point when the work item of interest, i.e., the diff, did not even exist yet.



\section{A System for Diff Prediction}
\label{S:facebook}

We will now sketch a system for work-item prediction which we
developed and deployed at Facebook. The system is focused on diffs as
work items, but it incorporates other types of work items, as a matter
of design~--~notably tasks (i.e., work items for project management
and planning at Facebook). The system leverages heuristics and machine
learning.

While the system can be used, for example, to address the `Aggregate
Performance' scenario of Section~\ref{S:scenarios}, it was initially
developed and deployed to address a specific security-related use
case, which is not discussed here for non-disclosure reasons.


\subsection{The Prediction Architecture}

The system essentially relies on a logging foundation that integrates
data for employee activity, a notion of time windows for work-item
prediction, a notion of candidate work-item IDs from which to pick,
and, ultimately, a ranking process for candidate work items per
employes.


\subsubsection{Logging Foundation}

Diff prediction relies on data from a number of logs; the most
important ones are these:

\begin{itemize}

\item Interactive, web-based tools~--~some of these tools cover
  development directly (e.g., the Phabricator UI, see below), but there
  are hundreds of tools, which are logged generically.
  
\item Version control (Mercurial, Git, etc.)~--~Checkouts are
  performed; commits are created, amended, and rebased. These actions
  are logged from the command line and the IDEs.

\item Code reviewing and continuous integration (Phabricator UI and CLI et
  al.)~--~Commits are published for review, tested in a sandbox,
  commented on, revised, accepted or rejected, landed in production or
  abandoned, etc.
  
\item Task management~--~Tasks are created, assigned,
  commented on, associated with diffs, closed, etc.

\item Development tools~--~Build project; run test; debug code; query
  a database; etc.
  
\end{itemize}


\subsubsection{Time Windows Into Dark Matter}

We aim at probabilities for a certain employee to work on a certain
work item (diff) at a certain time. We use time windows of 10 minutes
as the granularity on the time axis for aggregating signal and
learning correlations. For each employee (engineer), we use basic
logging data to determine the windows during which the employee was
active and for which prediction is thus relevant;
see~\autoref{L:active_time_windows}.


\begin{lstlisting}[
  language=Presto,
  float=h,
  caption=\pseudo{Active Time Windows},
  label=L:active_time_windows]
CREATE TABLE active_time_windows (
   employee BIGINT COMMENT 'Employee ID',
   first_time BIGINT COMMENT 'Window first time',
   last_time BIGINT COMMENT 'Window last time'
)
PARTITIONED BY (ds STRING) -- YYYY-MM-DD - for the day of the data
\end{lstlisting}


Throughout this section, we use such relational table schemas to hint
at the data model used by the system for diff prediction, which is
implemented in Facebook's data warehouse while relying on Hive, Spark,
Presto, and scheduled pipeline and ML runs. All tables are
partitioned by day (see `ds'), i.e., work-item prediction is is
generally approached on a per-day basis.

The time windows associate with corresponding event sequences as of
the logging foundation. (Think of join conditions based on time
boundaries.)  We also refer to the time-windowed event sequences as
`dark-matter sequences'; see the introduction for our use of the term
`dark matter'.


\subsubsection{Candidate Work Items}

Prediction cannot make up work items (or IDs thereof) by
itself. Instead, prediction is given access to a set of candidate work
items that can be at all expected to be worked on during a given day
by a given person.

To this end, we use high-confidence signal such as version control-
and reviewing-based interaction of employees with diffs such as the
events illustrated in~\autoref{F:diff-analyzer}. We propagate backward
$b$ days and forward $f$ days. (Let's assume here $f=b=2$.)  Thus, we
end up with a set of candidate diffs as modelled
in~\autoref{L:candidate_diffs}.


\begin{lstlisting}[
  language=Presto,
  float=h,
  caption=\pseudo{Candidate Diffs per Employee},
  label=L:candidate_diffs]
CREATE TABLE candidate_diffs (
   employee BIGINT COMMENT 'Employee ID',
   diff_number BIGINT COMMENT 'Candidate diff'
)
PARTITIONED BY (ds STRING)
\end{lstlisting}


\subsubsection{Prediction by Ranking}

On a given day, for each employee, for each active time window, and
for each candidate work item, we need to determine the probability of
the employee working during the time window on the work item. Many of
these probabilities should be expected to be zero. All these
probabilities, when combined, define directly a ranking of work items
per time window of the employee. The actual prediction is determined
by a combination of heuristics, applied to the dark-matter sequences,
subject to heuristic averaging. Some of the heuristics use machine
learning.

When outputting a prediction, we also record the contribution of each
heuristic, thereby contributing to an explainable model for work-item
prediction~\cite{Rudin19}.  The output format for predictions is
modelled in~\autoref{L:diff_predictions}.


\begin{lstlisting}[
  language=Presto,
  float=h,
  caption=\pseudo{Diff Predictions},
  label=L:diff_predictions]
CREATE TABLE diff_predictions (
   employee BIGINT COMMENT 'Employee ID',
   first_time BIGINT COMMENT 'Sequence first time',
   last_time BIGINT COMMENT 'Sequence last time',
   diff_number BIGINT COMMENT 'Candidate diff',
   prediction DOUBLE COMMENT 'Probability of employee working on diff',
   contributions MAP<STRING, DOUBLE> COMMENT 'Contributions of heuristics'
)
PARTITIONED BY (ds STRING)
\end{lstlisting}


The individual heuristics compute either a Boolean value or a
probabilistic value in the range $[0,1]$. Thus, there is a table of
per-heuristic output, subject to additional partitioning as modelled
in~\autoref{L:diff_heuristics}.


\begin{lstlisting}[
  language=Presto,
  float=h,
  caption=\pseudo{Diff Heuristics},
  label=L:diff_heuristics]
CREATE TABLE diff_heuristics (
   employee BIGINT COMMENT 'Sequence employee ID',
   first_time BIGINT COMMENT 'Sequence first time',
   last_time BIGINT COMMENT 'Sequence last time',
   diff_number BIGINT COMMENT 'Candidate diff',
   label DOUBLE COMMENT 'Positive (1.0) / negative (0.0) label'
)
PARTITIONED BY (
   ds STRING,
   heuristic STRING -- Name of heuristic
)
\end{lstlisting}

The extra partitioning simplifies the process of computing the
individual heuristics in the data warehouse in a distributed
manner. (We could also use one table for each heuristic instead.) 


\subsection{Selected Heuristics}

Let us sketch some of the heuristics used by the system for diff prediction.


\subsubsection{Heuristic `\emph{Diff Analysis}' (DA)}
\label{SS:da}

We extract `strong confidence' events for authors and
reviewers interacting with diffs. All those events are labeled with
`1.0'. There are `obvious' events when diff authors submit diffs
or new versions thereof for reviewing or reviewers submit reviews or
parts thereof. We also incorporate so-called
work-in-progress events which are about an employee's interaction with
the repository and local commits~---~also before an actual Phabricator
diff is created. Consider the following sample workflow of an employee:

\begin{enumerate}
\item Update to master on the developer's machine.
\item Start editing and commit locally.
\item Continue editing and amend locally.
\item Split the amended commit.
\item Submit a stack of two diffs for review.
\end{enumerate}

At point (1), a checkout identifier (also referred to as
`work-in-progress' identifier) is created. Each of the operations
(2)--(4) creates more checkout identifiers and also commit hashes as
the result of the mutations. Once operation (5) creates two diffs,
actual diff IDs become available. We can now travel into the past and
connect timestamps (1)--(4), in this case, with both diffs. This
heuristic leverages an integrated event log for version control
and code review, as modelled in~\autoref{L:diff_event_log}.

\begin{lstlisting}[
  float=h,
  language=Presto,
  caption=\pseudo{Integrated Event Log for Diffs},
  label=L:diff_event_log]
CREATE TABLE diff_event_log (
   id BIGINT COMMENT 'Diff event ID',
   time_started BIGINT COMMENT 'Time the event (action) started',
   time_ended BIGINT COMMENT 'Time the event (action) ended',
   actor BIGINT COMMENT 'ID of employee acting on the diff',
   event_type STRING COMMENT 'Type of diff event',
   diff_number BIGINT COMMENT 'Diff number',
   version_number BIGINT COMMENT 'Version number of diff',
   owner BIGINT COMMENT 'ID of employee owning the diff',
   data STRING COMMENT 'Extra metadata in JSON'
)
PARTITIONED BY (ds STRING)
\end{lstlisting}


\subsubsection{Heuristic `\emph{Task Events}' (TE)}

Tasks support project management and planning at Facebook.  There are
tasks for features to be developed, incidents to be investigated, bugs
to be fixed, etc.  Diffs are typically also associated with tasks
eventually. Once this association is revealed, we count task
interaction events retroactively towards associated diffs.


\subsubsection{Heuristic `\emph{Diff URIs}' (DU)}
\label{SS:du}

The web-based, internal tools used at Facebook may track work items,
to some extent, through URI parameters. For instance, diff IDs are
generally represented in this format ``\verb!D(\\d+)!'' (using regular
expression syntax here) and this representation is also used within
URIs. There is a few specific tools, for which we are readily aware of
their usage of diff IDs in the URIs and specific positions or
parameters thereof. The present heuristic extracts diff-ID occurrences
more generically. Without a heuristic like this, we would need to
continuously invest into extraction functionality for a multitude of
evolving and new tools. The heuristic is prone to false positives, but
aims at minimizing those by incorporating knowledge of the valid
length of contemporary diff IDs. (For instance, the string `D42' would
be considered too short to count as a diff ID.)


\subsubsection{Heuristic `\emph{Diff Comparison}' (DC)}

Based on backup system-like support for tracking local repository
changes, we track `file changes', ahead of actual commits so that they
can be correlated with diffs, eventually.

We tokenize (featurize) diffs and changes:

\begin{itemize}
\item filename extensions;
\item filename words (directory names, basenames, parts thereof);
\item symbols `used' (e.g., program identifiers referenced).
\end{itemize}

For instance, filename extensions help to already group diffs largely
by language (technology); filename words help with aligning changes
based on the affected regions in the file tree. We leverage similarity metrics
(cosine et al.) and clustering for comparison, in fact, similarity
analysis.


\subsubsection{Heuristic `\emph{Word Indexing}' (IX)}
\label{SS:ix}

The `Diff URIs' (DU) heuristic suffices when diff IDs are explicitly
mentioned in URIs. We devise the `Word Indexing' (IX) heuristic to
better address the defining characteristics of the WIP challenge to
involve a multitude of loosely integrated tools along concurrent,
long-running workflows.

Consider the following (obfuscated) URI, which we keep trivial for
ease of explanation, but it should be noted that URIs for internal
tool usage at Facebook can be rather complex, as significant
session/context data is captured by the URIs:

\smallskip

{\footnotesize
  
\begin{center}
  \verb!https://internal.acme.com/tools/tool-42/resources/123/?task=T4711!
\end{center}

}

\smallskip

The heuristic relies on tokenization of URIs. That is, the following
words would be extracted for the example: tool, 42, resources, 123,
task, T4711. The heuristic performs word indexing and, in fact,
word-correlation learning such that, for example, we may infer that
`T4711' associates generally (probabilistically) with a certain diff
ID, if `T4711' co-occurs with the diff ID in enough time windows
elsewhere.  This is explained in Section~\ref{S:ix} in more detail.


\subsubsection{Heuristic Averaging}

The heuristics are combined to compute a final prediction by weighted
averaging. We use larger weights for higher confidence signal (such as
the DA heuristic) and lower weights for lower confidence signal, to
account for noise (false positives), subject to the overall assumption
that low weights are still sufficient to create a rank for work items
and a combination of low weights may also increase the rank.


\begin{figure}[t!]

 {

\centering

\includegraphics[width=.45\columnwidth]{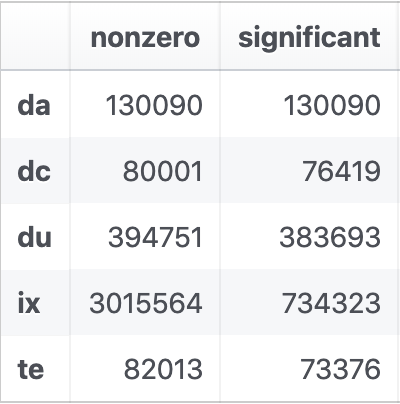}

}
\vspace{-22\in}
\caption{A Sample of Two Basic Heuristics Metrics.}
\label{F:heuristics-metrics}
\vspace{-42\in}
\end{figure} 


\autoref{F:heuristics-metrics} shows two basic metrics for the
selected heuristics, as they affect the final predictions~--~this is a
sample for a recent day and some group of employees. The
\emph{nonzero} column counts how many time windows are labeled by the
heuristic. The \emph{significant} column counts how many times a
significance threshold is passed, subject to internal validation for
reducing false positives. We mention in passing that we use yet other
metrics, for example, a quantification of whether a work item is
included exclusively due to a specific heuristic or a distribution of
the distance between high-confidence signal (DA and TE) and
low-confidence signal (DC and IX) per pairs of work item and employee.


\subsection{Word-Correlation Learning}
\label{S:ix}

Let us discuss the IX heuristic for word-correlation learning, as
introduced in Section~\ref{SS:ix}, in more detail, as this component
addresses the defining characteristics of the WIP challenge in a
relatively advanced manner.


\subsubsection{Tokenization}

The tokens (words) extracted from URIs due to tool usage are the
starting point for word indexing; see~\autoref{L:uri_words}.


\begin{lstlisting}[
  float=h,
  language=Presto,
  caption=\pseudo{Words in URIs due to Tool Usage},
  label=L:uri_words]
CREATE TABLE uri_words (
   employee BIGINT COMMENT 'Employee ID for tool usage',
   time BIGINT COMMENT 'Time the tool was used',
   uri_words ARRAY(STRING) COMMENT 'Words in a URI for tool use'
)
PARTITIONED BY (ds STRING)
\end{lstlisting}


\subsubsection{Word Frequency}

Many of the URI words (tokens) extracted from the URIs for tool usage
are `noise' in the sense that they occur all
too often and we need to filter the set of words to ever be considered
for prediction. To this end, we aggregate, over time, the frequency of
words; see~\autoref{L:word_frequency}.


\begin{lstlisting}[
  float=h,
  language=Presto,
  caption=\pseudo{Word Frequency},
  label=L:word_frequency]
CREATE TABLE word_frequency (
   uri_word STRING COMMENT 'Word in a URI for tool use',
   word_days BIGINT COMMENT 'Number of employee days the word appears in',
   all_days BIGINT COMMENT 'Total number of available employee days',
   inverse_frequency DOUBLE COMMENT 'IDF computed from above numbers'
)
PARTITIONED BY (ds STRING)
\end{lstlisting}


In particular, we compute an inverse document (-like) frequency with
all documents corresponding to the days seen by our analysis versus
those days where the word occurs. We omit here details how exactly we
apply filters, but we use this frequency table in an obvious sense to
focus on words of interest and to maintain scalability of the
prediction process. Here we note that several 100K new words show up
every day in our dataset.


\subsubsection{Co-Occurrence with Diff IDs}

We determine co-occurrences of diff IDs for candidate diffs with other
words within URIs. We build a corresponding index of such word
overlaps per day from a number of past days and we track the
employee for whose dark-matter sequence the co-occurrence
occurred; see~\autoref{L:diff_id_overlaps}.


\begin{lstlisting}[
  float=h,
  language=Presto,
  caption=\pseudo{Overlaps of diff IDs and Other Words},
  label=L:diff_id_overlaps]
CREATE TABLE diff_id_overlaps (
   employee BIGINT COMMENT 'Employee ID',
   first_time BIGINT COMMENT 'Sequence first time',
   last_time BIGINT COMMENT 'Sequence last time',
   diff_number BIGINT COMMENT 'Candidate diff',
   word STRING COMMENT 'Word that co-occurs',
   index_employee BIGINT COMMENT 'Employ with overlap in index',
   index_time BIGINT COMMENT 'Overlap time in index',
   index_ds STRING COMMENT 'Overlap day in index'
)
PARTITIONED BY (ds STRING)
\end{lstlisting}


\subsubsection{Word-Indexing Features}

At this point, we can extract features from word overlaps;
see~\autoref{L:word_index_features} for some examples of count-based
features.


\begin{lstlisting}[
  float=h,
  language=Presto,
  caption=\pseudo{Word Index Features},
  label=L:word_index_features]
CREATE TABLE word_index_features (
    employee BIGINT COMMENT 'Employee ID',
    first_time BIGINT COMMENT 'Sequence first time',
    last_time BIGINT COMMENT 'Sequence last time',
    diff_number BIGINT COMMENT 'Candidate diff',
    ft_ct_total_overlaps INT COMMENT
        'Total number of overlaps of diff ID with other words.'
    ft_ct_overlaps_more_distinct INT COMMENT
        'Number of distinct triples (word, employee, ds) from among
        all the overlaps. In particular, this does not count repeated visits 
        to the same word within the sequence.',
    ft_ct_overlapping_words INT COMMENT
        'Number of distinct words from among all the overlaps. This counts 
        how many related words there are, without looking at how closely 
        related each individual word is.',
    ft_ct_overlapping_employee_days INT COMMENT
        'Number of distinct employee/ds pairs from among all the overlaps.
        This counts how many times the sequence and candidate can be related,
        without looking at how many words they are related by each time.',
    ft_min_overlap_employee_day_freq INT COMMENT
        'Smallest employee-day frequency of an overlapping sequence word. 
        That is, from all the overlaps, find the sequence word that is rarest 
        in general (by employee-day frequency, i.e., by counting the number of 
        employee/ds pairs in which it appears in the past).',
    ...
)
PARTITIONED BY (ds STRING)
\end{lstlisting}


For each feature, one may have an intuition as to why the feature
could be possibly helpful for work-item prediction. For instance,
\emph{ft\_min\_overlap\_employee\_day\_freq} is included because
overlapping words that are rarer in the `entire' past of an employee
may be more meaningful.


\subsubsection{Prediction with Decision Trees}
\label{SS:ml}

We leverage a decision tree (DT) algorithm, in fact, GBDT with
permutation-based feature importance. We rely on a weak supervision
approach as follows. For \emph{positive labeling}, we use
high-confidence events. For \emph{negative labeling}, essentially, we
select dark-matter sequences without high-confidence events, but
instead with high confidence of no diff-related work based on an
observable, prolonged hiatus regarding diffs in terms of absence of
high-confidence events, despite though being active overall.


\begin{figure}[t!]

\vspace{-42\in}
  
 {

\centering

\includegraphics[width=1\columnwidth]{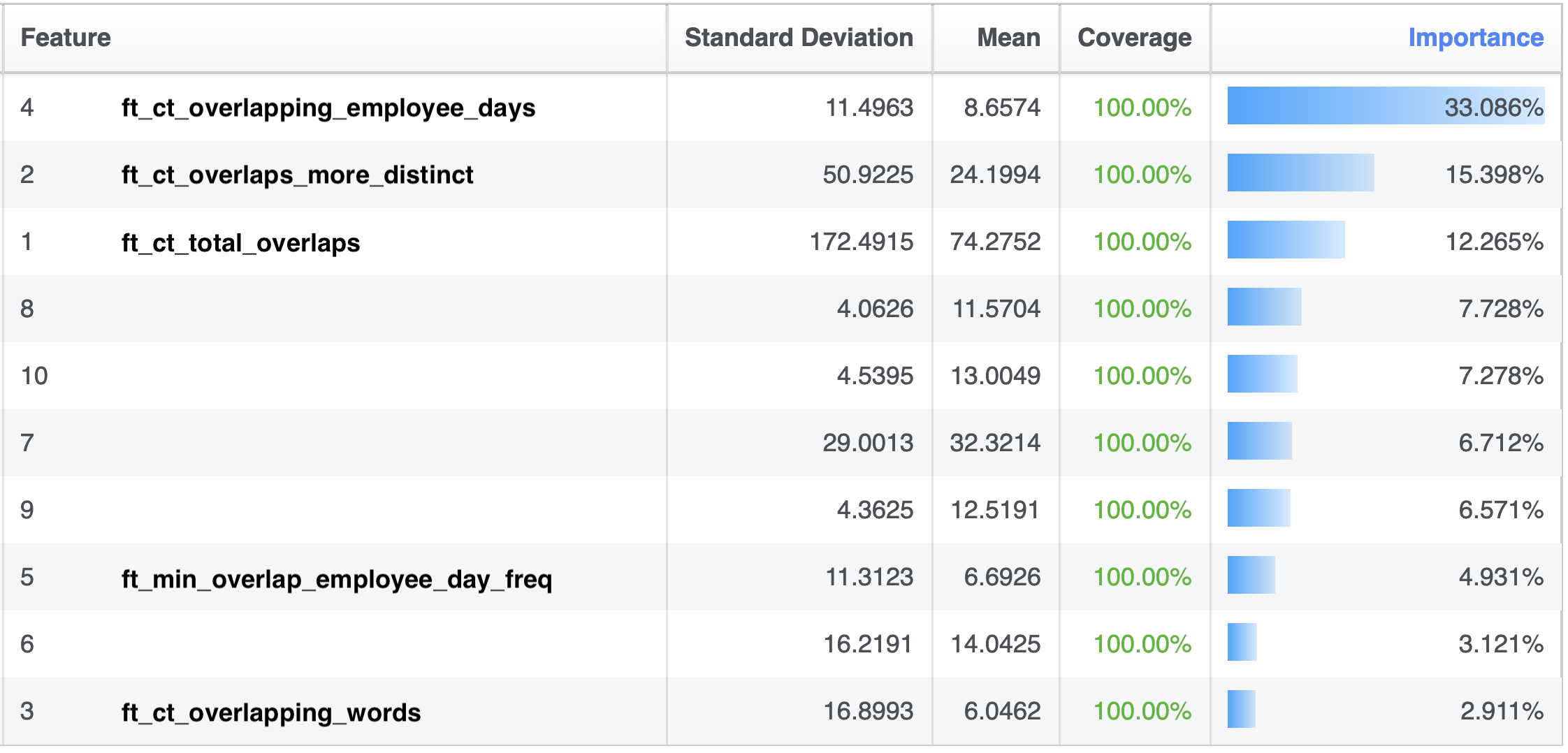}

}

\vspace{-22\in}

\caption{Feature Importance of the GBDT Model.}
\label{F:feature-importance}
\vspace{-22\in}
\end{figure} 


In~\autoref{F:feature-importance}, we illustrate feature
importance. For instance, the importance of
\emph{ft\_ct\_overlapping\_employee\_days} is evident, while our hopes
for \emph{ft\_min\_overlap\_employee\_day\_freq} are
shattered. (Disclaimer: feature importance is shown here at some point
during development, feature engineering, and feature selection.)



\section{Lessons Learned}
\label{S:lessons}

Based on our work on work-item prediction at Facebook (diffs and other types of work items), we share some lessons learned. For non-disclosure reasons, we cannot report empirical results on work-item prediction. Instead, we focus here on more general findings.
 

\subsection{The Dominance of Event Log Integration}

Events of interest are scattered over diverse available logs. Thus, filtering, abstraction, cleaning, alignment, and quality checking efforts are crucial. In our experience, this work combined with recovering the semantics of the given data dominates efforts on work-item prediction. Some of these integration efforts can be technically involved and require significant computational and storage investments. For instance, diff prediction required an extra component to perform chasing from diffs to local commits based on version control data~---~this turned out to be a non-trivial graph-based computation.


\subsection{The Effort of Managing Prediction DAGs}

We implement prediction architectures as directed acyclic graphs (DAG) with tables as nodes and scheduled pipeline and ML runs in the data warehouse as edges. For instance, the DAG for diff prediction includes many dozens of tables and scheduled runs. (Arguably, splitting up data into tables and computations into scheduled runs is subject to some design choices. Also, some of the tables are reused or reusable.) We suffice with offline, day-by-day prediction and thus, all computation is scheduled for daily execution.

The challenge is that a prediction architecture has significant internal and upstream dependencies; we are affected by delays, failures, or changes in the data semantics. It requires continuous effort and rigorous engineering (e.g., monitoring, data-quality checking, responding to alerts) to keep the overall DAG up and running.


\subsection{Limits of Human-Based Validation}

A naive approach to human-based validation would require subjects to accept or reject triples of the form time window, employee, and work item, thereby providing labeled data for machine learning and enabling measurements of model performance (e.g., precision and recall) in the most direct manner. The costs of such an approach are arguably prohibitive in an industrial setting. Also, it is relatively hard to sufficiently help the subjects to make an informed decision. There are more indirect means of human-based validation, which also can be used for weak supervision. i) Gamification (For instance, 'Hello X, were you working on diff D123, when you were running the database query Q456 at 2:42pm yesterday? Please, answer with Yes or No.') ii)
Efficiency (Can we measure that, for example, the incident response time is lower, on average, when responders use predicted work items as opposed to using more basic means of identifying work items based on, for example, simple queries for high-confidence events?) iii) Relevance (Can we measure how often an incident responder relies on predicted work items?)


\subsection{The Need for Internal Validation}

Especially because human-based validation is so limited, we need to use internal validation. To this end, some quantitative means are used: i) Predictions associate a significant number of time windows with specific work items. ii) A significant number of the time windows in i) do not contain high-confidence events, but only low-confidence events for the some work item. iii) With increasing life time of a work item (say, the span between first and last high-confidence event), the probability increases that there are interruptions in time windows being associated with the work item. These forms help in establishing basic properties of prediction. That is, prediction finds work items, the findings are not just based on proximity, and association of time windows with work items is on and off over time.


\subsection{The Value of Similarity Measures}

When we started our efforts on work-item prediction, we were enthusiastic about clustering work items (diffs) to correlate events (such as file changes or database queries, debugging sessions, or code searches) with eventual diffs. Perhaps obviously, clustering turned out to be the wrong tool for the problem because any discoverable clustering semantics within developer `utterances' turns out to be too coarse-grained to be useful in distinguishing, for example, diffs of the same developer, especially when clustering is performed for all developers at once. Also, the size of many utterances is just too small to be useful for clustering; consider, for example, short debugging sessions or trivial database queries. It turned out that relatively simple similarity measures (e.g., cosine) are more effective for correlating `utterances' with diffs.


\subsection{Systematic Dark Matter Elimination}

In our work, we included additional heuristics based on an ad-hoc process. In the case of diff prediction, we started from obvious, high-confidence events (such as `publish a diff'); we advanced towards inclusion of correlated events (such as `commit a version locally'), subject to some data integration; eventually, we leveraged some data mining and machine learning components. While the inclusion of any heuristic was driven by domain knowledge, the process is prone to premature optimization and gaps or delays regarding (internal) validation. We are now looking for a more data science-driven approach, where each decision to include an additional or to advance an existing heuristic can be supported by an appropriate analysis.



\section{Related Work}
\label{S:related}

Let us review related to work so that we connect the WIP challenge and the sketched efforts at Facebook to prior art. Related work is from the areas of program comprehension, mining software repositories, process mining, and machine learning. We group related work entries by some common themes.

\newcommand{\densepara}[1]{\paragraph*{#1} \vspace{-27\in}}


\subsection{Developer Workflow Mining}

\paragraph*{\textbf{Summary}} Work in this group tends to involve explicit models of developer workflow; making such an assumption would make the WIP challenge only harder. None of the scenarios that we have in mind require an explicit model. Further, some of the work in this group touches upon the issue of using multiple tools, but the more general characteristics of a multitude of loosely integrated tools along concurrent, long-running workflows are not addressed. However, work in this group submits important techniques for event modeling and data mining, overall.

\densepara{Visualizing IDE sessions~\cite{MinelliL13}} The approach is inspiring in terms of setting up some basic concepts such as IDE sessions (to provide scope), events and classifying them more abstractly (e.g., inspection, editing, and navigation), and navigation paths to express that different entities are manipulated. The frequency of events and navigation paths are visualized by size/width. Colors are used to encode classification properties.

\densepara{Process mining for mining software repositories~\cite{PoncinSB11}} In one of the case studies, the paper uses the ProM tool for process mining to extract the bug life cycle in a software project. Much of the work is concerned with preprocessing data sources for use with the ProM tool for process mining.

\densepara{Identification of usage smells in IDEs~\cite{DamevskiSSP17}} This work helps assessing the usability of an IDE. Event sequences are analyzed through stages of pattern mining (mostly counting common sequences with some degree of variation), pattern filtering (e.g., to remove too short sequences), and pattern clustering (to compress the large number of patterns to be amenable to visual inspection).

\densepara{Workflow mining from IDE usage~\cite{IoannouBW18}} The Disco tool is used to extract the actual workflows in terms of graphs with events and frequency-annotated transitions. Events are classified in some interesting ways (e.g., discrete versus continuous events). Workflows are determined in an experiment-based manner such that developer subjects solve certain tasks with the given IDE.

\densepara{Prediction or recommendation of developer behavior in the IDE~\cite{DamevskiCSKP18}} Temporal Latent Dirichlet Allocation (Temporal LDA) for topic modeling is applied to IDE interaction data. In this manner, high-level task behavior such as structured navigation is discovered based on interpreting sequences of lower level interaction events and commands.

\densepara{Developer behavior across GitHub and StackOverflow~\cite{XiongMSY17}} The accounts are associated and cross-site developer behaviors are analyzed through T-graph analysis, topic-based clustering (based on LDA) and cross-site tagging. 

\densepara{Discovering software processes from OSSD Web repositories~\cite{JensenS04}} A general methodology is presented, combining techniques for text analysis, link analysis, and patterns of repository usage and update. This work presents a process entity taxonomy and relies on the construction of a social graph to better capture who is working with whom how. Probabilistic relational modeling (PRM) is used to model what the developers are doing and how they are doing it.

\densepara{Correlation of code changes with reviewer input~\cite{BellerBZJ14}} This correlation is measured while assuming a certain process for code review with involvement of a task management system. The work does not use any advanced information retrieval techniques, but it relies instead on carefully collected datasets and subject-based ground truth. A significant percentage of code changes is classified as not responding to reviewer comments.


\subsection{Case ID Recovery in Process Mining}

\paragraph*{\textbf{Summary}} Work in this group applies to \emph{process-unaware systems}~\cite{GoelBW13} in so far that case IDs (say, work-item IDs~--~for our purposes) are not assumed. However, other characteristics of the WIP challenge (the multitude of evolving tools, the significance of concurrent workflows, the limitations of logging) are not taken into account.

\densepara{Key alignment by composite key conditions~\cite{NezhadBSCA08}} The work aims at integrating separate logs using different kinds of keys. Alignment of the keys relies on composite key conditions of some specific format, which assumes key attributes, equality, conjunction, and disjunction, so that entries from different logs are assigned to the same workflow instance, when the conditions hold. The conditions do not need to be designed manually, but they can be explored semi-automatically by a heuristic, thereby providing a form of case ID learning.

\densepara{Case ID assignment based time proximity~\cite{FerreiraG09} et al.} The probability of certain events to follow certain other events can be inferred from an (unlabeled) log. Once such a probability matrix has been estimated, assignment of case IDs is essentially an optimization problem. This approach is of limited use in a setting, like the one of the WIP challenge, with concurrent and long-running workflows and an event log that necessarily contains much `noise' in terms of irrelevant events.

\densepara{Case IDs in a mobile app context~\cite{DuEtAl17}} The service log, which logs requests of the app, is translated to an event log while filling in missing case IDs on the grounds of several heuristics: some form of time proximity, spatial continuity, grouping of requests by source, and an overall time bound (due to app specifics). The work caters for multiple versions of the underlying process, thereby addressing system evolution to some extent.


\subsection{Machine Learning in Process Mining}

\paragraph*{\textbf{Summary}} Work in this group discusses machine learning in the process mining context. The work does not directly address the defining characteristics of the WIP challenge, but the work inspires potential advances of any WIP solution to better address event abstractions, ML concept drifts, and representations for embeddings, or to enable additional applications such as anomaly detection.

\densepara{Event abstraction~\cite{TaxSHA16b}} Supervised learning is leveraged for event abstraction based on annotations with high-level interpretations of low-level events while relying on extensions of the XES event-stream format and using conditional random fields for machine learning. (Without event abstraction, raw events may be too fine-grained for process discovery to return useful results.)

\densepara{Discovering and understanding potential concept drifts~\cite{BoseAZP14}} (This term is used in machine learning to refer to situations when the relation between the input data and the target variable, which the model is trying to predict, changes over time in unforeseen ways, thereby letting accuracy of the predictions degrade over time.) This work provides a generic framework and specific techniques for discovering and understanding potential concept drifts in process mining. In particular, the effects of  'noise' (random replacements or insertions of incorrect symbols or missing symbols) and imbalance (largely different priorities of certain branches) are addressed.

\densepara{Representation learning~\cite{KoninckBW18}} The work is based on the observation that, we quote, ``real-life event logs present a large number of cases, potentially representing a highly varied set of distinct event sequences, and usually also containing information on resources and a diverse set of other event- or case-related attributes''. As a consequence, featurized event data suffers from a dimensionality problem, which the paper addresses by representation learning architectures for activities, traces, logs, and process models.

\densepara{Anomaly detection~\cite{NolleSM18}} Neural networks are applied to address a general objective in process mining: verification of the absence of certain undesirable properties or the compliance with certain desirable properties~\cite{AalstBD05,AccorsiSM13,FerilliE13}.



\section{Concluding remarks}
\label{S:concl}

In this paper, we have established the work-item prediction challenge, as it applies to software engineering workflows in software-intensive organizations.

The defining characteristics of the challenge are these:
\defwip

In practice, an ensemble of heuristic- and ML-based components is needed to predict work items on the timelines of employees.  We have described related efforts at Facebook, where diffs (system changes), as a type of work item, are predicted. Such prediction is readily useful in the context of incident response in engineering (with relevance for reliability, integrity, privacy, and security) or the aggregation of key performance indicators for engineering processes.

We have provided an extended related work discussion which connects the work-item prediction challenge to research in the areas of program comprehension, mining software repositories, process mining, and machine learning. This discussion documents the need for and the potential of leveraging, advancing, and combining existing techniques to tackle the work-item prediction challenge more efficiently in practice.


\pagebreak

\nocite{Praza19}
\bibliography{paper}
\bibliographystyle{ACM-Reference-Format}

\end{document}